\documentclass[twocolumn,showpacs,prl]{revtex4}

\usepackage{graphicx}
\usepackage{dcolumn}
\usepackage{bm}

\usepackage{graphics}
\begin{document}

\title{The pulling force of a single DNA molecule condensed by spermidine}

\author{R. Zhang}
\author{B. I. Shklovskii}
\affiliation{Theoretical Physics Institute, University of
Minnesota, Minneapolis, Minnesota 55455}

\date{\today}

\begin{abstract}
In a recent experiment, a single DNA double helix is stretched and
relaxed in the presence of spermidine, a short positive
polyelectrolyte, and the pulling force is measured as a function
of DNA extension. In a certain range of spermidine concentration,
a force plateau appears whose value shows maximum as a function of
spermidine concentration. We present a quantitative theory of this
plateau force based on the theory of reentrant condensation and
derive almost parabolic behavior of the plateau force as a
function of the logarithm of the spermidine concentration in the
range of condensation. Our result is in good agreement with
experimental data.
\end{abstract}


\maketitle

DNA condensation by strongly positively charged proteins,
histones, is used by nature for compaction of DNA in cell nucleus
(DNA  is strongly negatively charged). Positive proteins,
protamines, are used for additional compaction of DNA in the
sperm~\cite{cell}. Gene therapy uses complexes of DNA with long
positive polyelectrolytes or other macrocations. The net charge of
these complexes can be positive and therefore they are not
repelled by negative cell membrane in the course of gene delivery.
Thus, condensation of DNA by positive macrocations is an extremely
important physical phenomenon. It is intensively studied in the
model system of double helix DNA with spermidine, a flexible
polymer with length $15$ $\AA$ and charge $+3$. It is known that
in a dilute DNA solution at some concentration of spermidine,
$s=s_c$, each long DNA molecule self-condenses into a
toroid~\cite{Bloomfield0}. When $s$ grows farther, at a much
larger concentration, $s=s_d$, DNA dissolves
back~\cite{Experiments}. This phenomenon is called reentrant
condensation and has got theoretical
explanations~\cite{Olvera,Rouzina}.
\begin{figure}[ht]
\begin{center}
\includegraphics[width=0.5\textwidth]{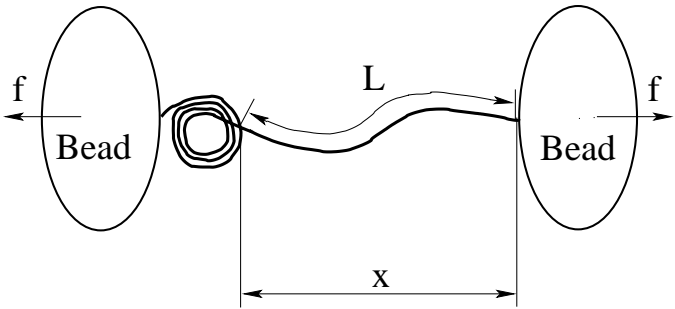}
\end{center}
\caption{Schematic illustration of the experimental setup. Partly
condensed long DNA is pulled by two beads.} \label{fig:coil}
\end{figure}

\begin{figure}[ht]
\begin{center}
\includegraphics[width=0.4\textwidth]{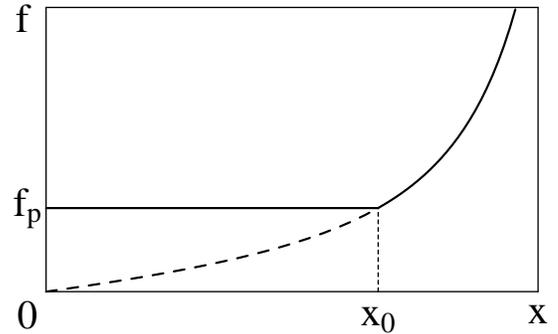}
\end{center}
\caption{The pulling force $f$ measured as a function of extension
$x$ is schematically shown by the solid line for $s_c<s<s_d$, when
DNA is condensed by spermidine. At $x<x_0$, the force $f_p$ is a
constant whose value depends on $s$. The dashed line shows the
wormlike chain behavior of $f$ in the absence of DNA condensation
($s<s_c$ or $s>s_d$). At $x>x_0$, the condensate is completely
eliminated by the pulling force and the wormlike chain behavior is
recovered.} \label{fig:plateau}
\end{figure}

\begin{figure}[ht]
\begin{center}
\includegraphics[width=0.5\textwidth]{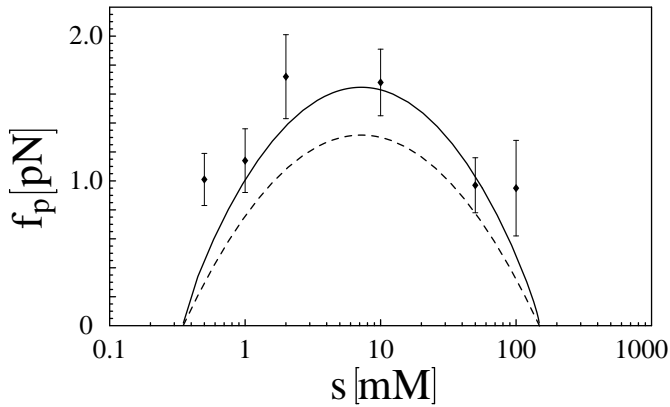}
\end{center}
\caption{Comparison between the theoretical result (the solid
line) and experimental data (points with error bars taken from
Fig.~2 of Ref.~\cite{Murayama}) for the pulling force $f_p$ as a
function of spermidine concentration $s$. $f_0$ is shown by the
dashed line.} \label{fig:ffd}
\end{figure}

Recently, a new single molecule technique has been used to measure
the force necessary to pull a single DNA double helix from a
toroidal condensate of DNA~\cite{Murayama} (Fig.~\ref{fig:coil}).
In the experiment, extremely small concentration of DNA (2 nM of
nucleotides) is dissolved in water with certain concentrations of
spermidine (varying from 200 $\mu$M to 200 mM). Dual-trap tweezers
are used to stretch a single DNA molecule tethered between two
protein-coated polystyrene beads. DNA is first stretched and then
relaxed. During these processes, the pulling force $f$ is
measured. Experiments are repeated at various spermidine
concentrations. It is observed that in the interval $s_c<s<s_d$,
there is a fairly large range of DNA extension, $x$, where force
$f$ is significantly larger than the force described in the
wormlike chain model~\cite{Marko}. Moreover, in this range of $x$,
$f$ is almost a constant (a force plateau, see
Fig.~\ref{fig:plateau}). The plateau value of the force, $f_p$, is
plotted as a function of $s$ in Fig.~\ref{fig:ffd}.

In this paper, we suggest a quantitative theory of the $f_p(s)$
curve based on the theory of reentrant
condensation~\cite{Rouzina}. We assume that the whole process is
slow enough so that the system is always in equilibrium. Then in
zero order approximation, the force needed to detach DNA from the
condensate is just the free energy difference per unit length
between the free coil state and the condensed state of the
DNA-spermidine complex. We call it $f_0$. The pulling force $f_p$
measured in the experiment is somewhat larger than $f_0$ since the
detached part of the DNA molecule is stretched, and therefore
loses entropy with respect of its free coil state. In the end of
the paper, we show that this effect adds a small correction to the
pulling force,
\begin{equation}
f_p=f_0\left(1+\sqrt{\frac{k_BT}{lf_0}}\right)\label{eq:fasym1}
\end{equation}
when $k_BT/lf_0 \ll 1$. Here $l$ is the persistence length of DNA.
In this paper, we first calculate $f_0$ and then $f_p$ using
Eq.~(\ref{eq:fasym1}). Our result is shown in Fig~\ref{fig:ffd}.
It demonstrates reasonably good agreement with experimental data.

Let us start from the theory of reentrant condensation of DNA with
spermidine without any external forces. In the solution,
positively charged spermidine molecules are absorbed on the
negatively charged surface of double helix DNA. Since spermidine
molecules are strongly charged ($+3$), they form a two dimensional
strongly correlated liquid on the surface of the DNA molecule.
When a new spermidine molecule approaches the DNA helix, it repels
already absorbed spermidine molecules and creates an electrostatic
image of itself, similar to the image on a conventional metallic
surface. Attraction to the image leads to an additional negative
chemical potential $\mu_c$ for spermidine molecules on the surface
of DNA. As a result, if $s$ exceeds some concentration, $s_0$,
charge of the DNA-spermidine complex changes sign and becomes
positive, i.e., spermidine molecules overcharge DNA (see review of
the theory of charge inversion in Ref.~\cite{Nguyen-review}).

Adsorbed layers of the correlated spermidine liquid also lead to
self-attraction of DNA. In the spot where two turns of DNA touch
each other, the surface density of the correlated liquid is
doubled and the correlation energy per spermidine molecule is
reduced~\cite{Bloomfield}. This short range force of
self-attraction leads to condensation of DNA if the macroscopic
Coulomb repulsion is not very strong. As a result, in the vicinity
of the neutralizing concentration $s_0$ ($s_c<s_0<s_d$),
DNA-spermidine complex is condensed into a toroid.

The condition of equilibrium between spermidine molecules in the
bulk and on the surface of DNA has a form
\begin{equation}
\mu_c+Ze\phi=k_BT\ln(sv_0),\label{eq:Zeq}
\end{equation}
where $Z=+3$ is the valence of spermidine, $\phi$ is the electric
potential on the surface of the DNA-spermidine complex, and $v_0$
is the normalizing volume of spermidine. The left hand side is the
chemical potential of a spermidine molecule in the complex, the
right hand side is its chemical potential in the bulk of the
solution. Notice that the entropy of spermidine can be expressed
through the total concentration $s$ because absorbtion of
spermidine by DNA practically does not change $s$ (the
concentration of DNA nucleotides is at least $10^{5}$ times
smaller than $s$). In the vicinity of $s_0$ ($s_c<s<s_d$), the net
charge of the complex is small and the spermidine concentration on
the DNA surface does not change much. Since $\mu_c$ is determined
by this concentration, it is approximately a constant. According
to Eq.~(\ref{eq:Zeq}), when $\phi=0$, i.e., the complex is
neutral, there is a simple relation between $\mu_c$ and $s_0$,
\begin{equation}
s_0=\frac{1}{v_0}\exp\left(-\frac{|\mu_c|}{k_BT}\right).
\label{eq:muc}
\end{equation}

Now we calculate the free energy of the charged complex in its
coil state. For this purpose, we treat the DNA-spermidine complex
as a capacitor with the capacitance per unit length $C$. Suppose
$s$ is such that the complex is overcharged ($s_0<s<s_d$).
Overcharging enhances the Coulomb self-energy and reduces both the
correlation energy and the entropy of spermidine molecules. Taking
the free energy of a neutral coil as zero, and using
Eq.~(\ref{eq:Zeq}), we get the free energy of the complex per unit
length
\begin{equation}
f_1=\frac{1}{2}C\phi^2+\frac{C\phi}{Ze}\mu_c-\frac{C\phi}{Ze}k_BT\ln(sv_0)
=-\frac{1}{2}C\phi^2,\label{eq:f1}
\end{equation}
where the three terms allow for the three parts of the free energy
mentioned above, and $C\phi/Ze$ is the number of spermidine
molecules overcharging the complex. The final expression is the
same as the free energy of a capacitor kept under a constant
voltage $\phi$. It is easy to see that this expression is also
true for an undercharged complex. Since the DNA-spermidine complex
can be considered as a long cylinder, $C$ is given by expression
\begin{equation}
C=\frac{D}{2\ln(1+r_s/R)},\label{eq:C}
\end{equation}
where $D=80$ is the dielectric constant of water, $r_s$ is the
Debye-H\"{u}ckel screening radius, and $R=10$ $\AA$ is the radius
of the double helix cross section.

We first assume that $r_s\gg R$. In this case, the complex in the
condensed state is practically neutral. Indeed, if the condensed
complex were charged, the Coulomb-self energy of the macroscopic
condensate would be too large to hold it. We define
phenomenological parameter $\varepsilon<0$ as the free energy per
unit length of the condensed complex calculated from the free
energy of a neutral coil. It includes the gain of correlation
energy in the spots where two turns of DNA touch each other, and
also the loss of certain entropic elasticity of the coil in the
condensate.

When the complex goes from the neutral condensed state to the
charged coil state, the free energy increment per unit length is
\begin{equation}
f_0=-\frac{1}{2}C\phi^2-\varepsilon.\label{eq:fd1}
\end{equation}

\begin{figure}[ht]
\begin{center}
\includegraphics[width=0.4\textwidth]{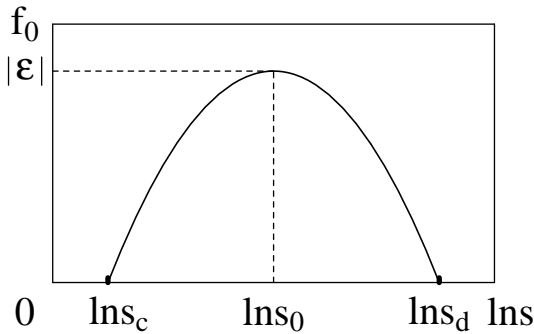}
\end{center}
\caption{The zero order approximation to the pulling force, $f_0$,
as a function of $\ln s$ given by an inverted parabola. The
maximum value of $f_0$, $|\varepsilon|$, is achieved at $s=s_0$
where the free DNA-spermidine complex is neutral. $s_{c,d}$ are
the concentrations of spermidine at which the condensate
dissolves.} \label{fig:tforce}
\end{figure}

If $r_s$ is comparable with $R$ (as in
experiment~\cite{Murayama}), the approximation used above that the
condensate is neutral has to be revised and the Coulomb energy of
the condensate should be taken into account. If we introduce
$C^{\prime}$ as the effective capacitance per unit length of the
complex in the condensate, similarly to Eq.~(\ref{eq:f1}), the
free energy per unit length is just $-C^{\prime}\phi^2/2$.
Accordingly, in Eq.~(\ref{eq:fd1}), $C$ should be replaced by
$C-C^{\prime}$. Using Eqs.~(\ref{eq:Zeq}) and (\ref{eq:muc}), we
rewrite Eq.~(\ref{eq:fd1}) as
\begin{equation}
f_0=|\varepsilon|-\frac{(C-C^{\prime})k_B^2T^2}{2Z^2e^2}\ln^2\frac{s}{s_0}.
\label{eq:F2}
\end{equation}
The function $f_0(\ln s)$ is shown in Fig.~\ref{fig:tforce} by an
inverted parabola. According to Eq.~(\ref{eq:F2}), the maximum
value $f_0=|\varepsilon|$ is achieved at $s=s_0$ where the free
complex is neutral. Also, we get the two concentrations $s_c$ and
$s_d$ at which $f_0$ goes to zero,
\begin{eqnarray}
s_{c,d}=s_0\exp\left(\mp\frac{1}{k_BT}\sqrt{\frac{2|\varepsilon|Z^2e^2}
{C-C^{\prime}}}\right), \label{eq:scd}
\end{eqnarray}
where the upper (lower) sign corresponds to the first (second)
subscript of $s_{c,d}$. Exactly at these two concentrations, one
can see transition from the coil state to the condensed state or
vice versa in light scattering experiments with solutions of DNA
and spermidine~\cite{Rouzina}.

In order to calculate $C^{\prime}$, we assume that the condensate
is macroscopic and densely packed. Then
\begin{equation}
\phi=\int_0^\infty\frac{\rho e^{-r/r_s}}{Dr}4\pi r^2dr=\frac{4\pi
r_s^2\rho}{D},
\end{equation}
where $\rho$ is the charge density of the condensate. The charge
of the complex per unit length is $\pi R^2\rho/\alpha$ where
$\alpha=0.91$ is the filling factor for the hexagonal dense
packing of cylinders. This gives
\begin{equation}
C^{\prime}=\frac{\rho}{\phi}\frac{\pi
R^2}{\alpha}=\frac{DR^2}{4\alpha r_s^2}.\label{eq:C'}
\end{equation}
We see that when $r_s\gg R$, we can drop $C^{\prime}$ in
$C-C^{\prime}$ because $C^{\prime}\ll C$, i.e., the condensate is
almost neutral.

As we mentioned before, $f_0$ is only the zero order approximation
to the pulling force. The pulling force $f_p$ is given by
Eq.~(\ref{eq:fasym1}). We need four experimental parameters to
calculate it. The spermidine concentrations $s_{c,d}$ at which the
condensate dissolves, the average DNA persistence length $l$, and
the average Debye-H\"{u}ckel screening radius $r_s$. Following
Ref.~\cite{Murayama}, we take $s_c=0.35$ mM, $s_d=150$ mM.
Therefore $s_0=7.2$ mM (Using Eq.~(\ref{eq:scd})). We use $l=500$
$\AA$ corresponding to the persistence length of a neutral DNA
coil. At $s_0=7.2$ mM, spermidine and its counter ion contribute
to $r_s$ even more than monovalent salt ($10$ mM). Treating every
spermidine molecule as a point-like trivalent ion, we get $r_s=13$
$\AA$ from the standard Debye-H\"{u}ckel expression~\cite{rs}.
Calculating $C$ and $C^{\prime}$ according to Eqs.~(\ref{eq:C})
and (\ref{eq:C'}), we finally get $|\varepsilon|=0.11$ $k_BT$/bp
(1bp $=3.4$ $\AA$) from Eq.~(\ref{eq:scd}).

For these parameters, $f_0$ and $f_p$ are shown together with the
experimental data in Fig.~\ref{fig:ffd}. We see that our result
for $f_p$ agrees pretty well with the experimental data. Notice
that correction $f_p-f_0$ is much smaller than $f_0$ in the range
of experimental data. This justifies the use of the perturbative
result Eq.~(\ref{eq:fasym1}).

Finally, let us derive Eq.~(\ref{eq:fasym1}). For this purpose, we
consider the detached part of DNA as a wormlike chain with the
contour length $L$ and end-to-end distance $x$ (see
Fig.~\ref{fig:coil}), while the contour length of the whole DNA
molecule is $L_0$. We also assume that the system is always in
equilibrium during the whole process. The free energy of the
system is
\begin{eqnarray}
F(x,L)=-(L_0-L)f_0+F_e \hspace{1.5in}\nonumber\\
=-(L_0-L)f_0+
\frac{k_BT}{l}\left[\frac{x^2}{2L}+\frac{L}{4(1-x/L)}-\frac{x+L}{4}\right],
\nonumber\\\label{eq:totalF}
\end{eqnarray}
where $-(L_0-L)f_0$ is the free energy of the condensed part of
DNA, and $F_e$ is related to the entropic elasticity of the
detached part of DNA. The expression for $F_e$ is obtained for a
\emph{free} wormlike chain with fixed contour length $L$ and
end-to-end distance $x$:
$F_e(x)=\int_0^{x}f_e(x^{\prime})dx^{\prime}+const.$. Here $f_e$
is the force needed to keep the end-to-end distance $x$ for a
\emph{free} wormlike chain with length $L$ and is given by an
interpolation formula~\cite{Marko}
\begin{equation}
f_e=\frac{k_BT}{l}\left[\frac{x}{L}-\frac{1}{4}+\frac{1}{4(1-x/L)^2}\right].
\label{eq:fe}
\end{equation}
The energy zero point has been chosen at $x=0, L=L_0$, i.e., at
the free energy of a DNA-spermidine complex in its coil state.

For a given $x$, the contour length $L$ of detached DNA can be
found from the condition of minimal free energy, namely, $\partial
F/\partial L|_x=0$. If we define $a\equiv x/L<1$, this minimum
condition can be written as
\begin{equation}
2a^2(1-a)^2+a^2-\frac{4lf_0}{k_BT}(1-a)^2=0.\label{eq:a}
\end{equation}
Finding $a$ from this equation and putting $L=x/a$ back to
Eq.~(\ref{eq:totalF}), we calculate the magnitude of the pulling
force
\begin{equation}
f_p\equiv\left|\frac{\partial F}{\partial x}\right|
=\frac{f_0}{a}+\frac{k_BT}{l}\left[\frac{a}{2}-\frac{1}{4}+\frac{1}{4(1-a)}\right].
\label{eq:f}
\end{equation}

Notice that $a$ does not depend on $x$ and $L$ (see
Eq.~(\ref{eq:a})). Therefore, according to the definition $x=aL$,
the two lengths $x$ and $L$ are always proportional to each other.
This result was also obtained in Ref.~\cite{Wada} using a
different approach. From Eq.~(\ref{eq:f}) it is easy to see that
$f_p>f_0$ since $a<1$. There are two contributions to $f_p-f_0$.
The first is a geometrical effect (the term $f_0/a$ in
Eq.~(\ref{eq:f})). Namely, to change the end-to-end distance by
$\Delta x$, a larger length $\Delta L=\Delta x/a$ must be pulled
out of the condensate. The second contribution is the free energy
increment of the detached part of DNA, $F_e$. These two
contributions are equal in the limiting case of $k_BT\ll lf_0$
(see Eqs.~(\ref{eq:fasym}), (\ref{eq:aasym}) and
(\ref{eq:fasym1})). Since $f_p$ does not depend on $x$, we get a
force plateau with increasing $x$, as observed in the experiment
(see Fig.~\ref{fig:plateau}). This plateau ends when $x$ reaches
$x_0=aL_0$ and all DNA is pulled out of the condensate. At
$x>x_0$, the pulling force starts to increase with $x$ according
to Eq.~(\ref{eq:fe})~\cite{ForcePlateau}.

When $k_BT/lf_0 \ll 1$, we have $1-a\ll 1$, and Eq.~(\ref{eq:f})
becomes
\begin{equation}
f_p=\frac{f_0}{a}+\frac{k_BT}{4l(1-a)}.\label{eq:fasym}
\end{equation}
Solving Eq.~(\ref{eq:a}) in the same limit, we get
\begin{equation}
a=1-\frac{1}{2}\sqrt{\frac{k_BT}{lf_0}}.\label{eq:aasym}
\end{equation}
Substituting Eq.~(\ref{eq:aasym}) in Eq.~(\ref{eq:fasym}), we
arrive at our final result Eq.~(\ref{eq:fasym1}). It gives the
pulling force $f_p$ in the first order perturbation theory in the
small parameter $\sqrt{k_BT/lf_0}$. In experiment~\cite{Murayama},
we have $k_BT/lf_0=0.06$ at $s=s_0$. Therefore,
Eq.~(\ref{eq:fasym1}) can be used in almost all the range
$s_c<s<s_d$. At $s$ close to $s_{c,d}$, where $f_0\rightarrow 0$,
the perturbation theory fails. In principle, we can calculate
$f_p$ numerically using Eqs.~(\ref{eq:a}) and (\ref{eq:f}) in the
whole interval $s_c<s<s_d$.

\begin{acknowledgments}
The authors are grateful to A. Yu. Grosberg, Y. Murayama, Y.
Rabin, I. Rouzina, and J. Zhang for useful discussions. This work
is supported by NSF No. DMR-9985785 and DMI-0210844.
\end{acknowledgments}



\begin{thebibliography}{100}
\bibitem{rs}This calculation overestimates the screening
effect because of the significant length of the spermidine
molecule. We can get a pretty good estimate of $r_s$ from above
considering each spermidine molecule as three separate positive
monovalent ions, which gives $r_s = 17$ $\AA$. Our final result is
not sensitive to this variation of $r_s$: the corresponding
variation of $|\varepsilon|$ is less than 7$\%$.
\bibitem{ForcePlateau} Strictly speaking, at $x\rightarrow x_0$,
our theory is not valid because the toroidal condensate becomes so
small that $f_0$ starts to decrease (the surface energy of the
toroid can not be ignored in this limit). Therefore at $x$ close
to $x_0$, the pulling force $f$ first decreases a little bit with
increasing $x$, and then goes back to the wormlike chain behaivor.
\end{thebibliography}
\end{document}